\documentclass{llncs}

\usepackage{amssymb}
\usepackage{amsmath}%
\usepackage{amsfonts}%
\usepackage{cmll} 
\usepackage{proof}
\usepackage{array}
\usepackage{enumitem}
\usepackage[table,x11names]{xcolor}
\usepackage{graphicx}
\usepackage{hyperref}

\usepackage[T1]{fontenc}

\usepackage{algpseudocode,algorithm,algorithmicx,float}
\algrenewcommand\alglinenumber[1]{\scriptsize #1:}

\newcommand{\quot}[1]{``\emph{#1}''}
\newcommand{\proc}[1]{\texttt{#1}}
\newcommand{\hl}[1]{\texttt{#1}}

\DeclareMathOperator{\tstile}{\vdash}
\DeclareMathOperator{\union}{\cup}
\DeclareMathOperator{\munion}{\uplus}

\newcommand{\mathhl}[1]{\mathtt{#1}}
\newcommand{\mathproc}[1]{\mathtt{#1}}
\newcommand{\memph}[1]{#1}

\newcommand{\sep}{,}
\newcommand{\inputtac}{$\mathsf{INPUT\_TAC}$}

\begin{document}

\title{A Pragmatic, Scalable Approach to Correct-by-construction Process Composition Using Classical Linear Logic Inference}
\titlerunning{Process Composition Using Linear Logic Inference}  

\author{Petros Papapanagiotou\inst{1} \and Jacques Fleuriot\inst{1}}
\authorrunning{Petros Papapanagiotou and Jacques Fleuriot} %

\tocauthor{Petros Papapanagiotou and Jacques Fleuriot}
\institute{School of Informatics, University of Edinburgh\\
10 Crichton Street, Edinburgh EH8 9AB, United Kingdom\\
\email{\{ppapapan,jdf\}@inf.ed.ac.uk}}

\maketitle              

\begin{abstract}
The need for rigorous process composition is encountered in many situations pertaining to the development and analysis of complex systems. We discuss the use of Classical Linear Logic (CLL) for correct-by-construction resource-based process composition, with guaranteed deadlock freedom, systematic resource accounting, and concurrent execution. We introduce algorithms to automate the necessary inference steps for binary compositions of processes in parallel, conditionally, and in sequence. We combine decision procedures and heuristics to achieve intuitive and practically useful compositions in an applied setting. 
\end{abstract}

\section{Introduction}
 
The ideas behind process modelling and composition are common across a variety of domains, including program synthesis, software architecture, multi-agent systems, web services, and business processes. Although the concept of a ``process'' takes a variety of names -- such as agent, role, action, activity, and service -- across these domains, in essence, it always captures the idea of an abstract, functional unit. Process composition then involves the combination and connection of these units to create systems that can perform more complex tasks. We typically call the resulting model a (\emph{process}) \emph{workflow}. Viewed from this standpoint, resource-based process composition then captures a structured model of the \emph{resource flow} across the components, focusing on the resources that are created, consumed, or passed from one process to another within the system.

Workflows have proven useful tools for the design and implementation of complex systems by providing a balance between an intuitive \textbf{abstract model}, typically in diagrammatic form, and a \textbf{concrete implementation} through process automation. Evidence can be found, for example, in the modelling of clinical care pathways where workflows can be both understandable by healthcare stakeholders and yet remain amenable to formal analysis~\cite{city1035,manataki2016}.
 
A scalable approach towards establishing trust in the correctness of the modelled system is that of \emph{correct-by-construction} engineering~\cite{KezadriHamiaz2016,cocsoa2013}. In general, this refers to the construction of systems in a way that guarantees correctness properties about them at design time. In this spirit, we have developed the \hl{WorkflowFM} system for correct-by-construction process composition~\cite{workflowfm}. It relies on Classical Linear Logic (see Section~\ref{sec:cll}) to rigorously compose abstract process specifications in a way that:
\begin{enumerate}%
\item systematically accounts for resources and exceptions;
\item prevents deadlocks; 
\item results in a concrete workflow where processes are executed concurrently. 
\end{enumerate}

From the specific point of view of program synthesis, these benefits can be interpreted as (1) \emph{no memory leaks or missing data}, (2) \emph{no deadlocks, hanging threads, or loops}, and (3) \emph{parallel, asynchronous (non-blocking) execution}.

The inference is performed within the proof assistant HOL Light, which offers systematic guarantees of correctness for every inference step~\cite{harrison1996hol}. The logical model can be translated through a process calculus to a concrete workflow implementation in a host programming language. 

There are numerous aspects to and components in the \hl{WorkflowFM} system, including, for instance, the diagrammatic interface (as shown in Fig.~\ref{fig:optexample}), the code translator, the execution engine, the process calculus correspondence, and the architecture that brings it all together~\cite{workflowfm}. In this particular paper we focus on the proof procedures that make such resource-based process compositions feasible and accessible. These are essential for creating meaningful workflow models with the correctness-by-construction properties highlighted above, but without the need for tedious manual CLL reasoning. Instead, the user can use high level composition actions triggered by simple, intuitive mouse gestures and without the need to understand the underlying proof, which is guaranteed to be correct thanks to the rigorous environment of HOL Light.

It is worth emphasizing that our work largely aims at tackling pragmatic challenges in real applications as opposed to establishing theoretical facts. We rely on existing formalisms, such as the proofs-as-processes theory described below, in our attempt to exploit its benefits in real world scenarios. As a result, the vast majority of our design decisions are driven by practical experience and the different cases we have encountered in our projects. 

Table \ref{tbl:projects} is a list of some of our case studies in the healthcare and manufacturing domain that have driven the development of \hl{WorkflowFM}. It includes an indication of the size of each case study based on (1) the number of (atomic) component processes, (2) the number of different types of resources involved in the inputs and outputs of the various processes (see Section~\ref{sec:spec}), (3) the number of binary composition actions performed to construct the workflows (see Section~\ref{sec:comp}), and (4) the total number of composed workflows.

\begin{table}[tpbh]
\centering
\rowcolors{2}{gray!25}{white}
\begin{tabular}{|l|c|c|c|c|}
\hline \textbf{Case study theme} & \textbf{Processes} & \textbf{Resource types} & \textbf{Actions} & \textbf{Workflows}\\ \hline\hline
Patient handovers & 9 & 16 & 13 & 2 \\
Tracheostomy care pathway & 33 & 47 & 32 & 3 \\
HIV care pathways & 128 & 129 & 121 & 13 \\
Pen manufacturing {\scriptsize\emph{(ongoing)}} & 42 & 45 & 60 & 20 \\\hline
\textbf{Total} & \textbf{212} & \textbf{237} & \textbf{226} & \textbf{38} \\\hline
\end{tabular}
\caption{Sample case studies and an indication of their size.}\label{tbl:projects}
\end{table}

All of these case studies are models of actual workflows, built based on data from real-world scenarios and input from domain experts such as clinical teams and managers of manufacturing facilities. The results have been useful towards process improvement in their respective organisations, including a better qualitative understanding based on the abstract model and quantitative analytics obtained from the concrete implementation. As a result, we are confident that the evidence and experience we accumulated from these case studies are representative of the requirements and needs of real applications and that the approach and algorithms presented in this paper can offer significant value.

We note that the presentation accompanying this paper is available online\footnote{\url{https://github.com/PetrosPapapa/Presentations/raw/master/LOPSTR2018.pdf}}.

\section{Background}

\begin{figure}[htbp]
 \centering
 \includegraphics[scale=10]{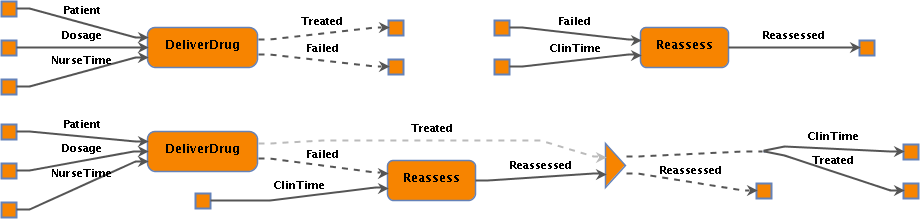}
 \caption{The visualisation of the \proc{DeliverDrug} and \proc{Reassess} processes (top) and their sequential composition. The auxiliary triangle helps properly display the output.}
 \label{fig:optexample}     
\end{figure}

The systematic accounting of resources in our approach can be demonstrated through a hypothetical example from the healthcare domain~\cite{workflowfm}. Assume a process \proc{DeliverDrug} that corresponds to the delivery of a drug to a patient by a nurse. Such a process requires information about the $Patient$, the $Dosage$ of the drug, and some reserved $NurseTime$ for the nurse to deliver the drug. The possible outcomes are that either the patient is $Treated$ or that the drug $Failed$. In the latter case, we would like to apply the \proc{Reassess} process, which, given some allocated clinician time ($ClinTime$) results in the patient being $Reassessed$. A graphical representation of these 2 processes, where dashed edges denote the optional outcomes of \proc{DeliverDrug}, is shown at the top of Fig.~\ref{fig:optexample}.

If we were to compose the 2 processes in a workflow where the drug failure is always handled by \proc{Reassess}, what would be the specification (or specifically the output) of the composite process?

Given the workflow representation in Fig.~\ref{fig:optexample}, one may be inclined to simply connect the $Failed$ edge of \proc{DeliverDrug} to the corresponding edge of \proc{Reassess}, leading to an overall output of either $Treated$ or $Reassessed$. However, this would be erroneous, as the input $ClinTime$, is consumed in the composite process even if \proc{Reassess} is never used. Using our CLL-based approach, the workflow output is either $Reassessed$ which occurs if the drug failed, or $Treated$ coupled with the unused $ClinTime$, as shown at the bottom of Fig.~\ref{fig:optexample} \cite{workflowfm}.

Systematically accounting for such unused resources is non-trivial, especially considering larger workflows with tens or hundreds of processes and many different outcomes. The CLL inference rules enforce this by default and the proof reflects the level of reasoning required to achieve this. In addition, the process code generated from this synthesis is fully asynchronous and deadlock-free, and relies on the existence of concrete implementations of \proc{DeliverDrug} and \proc{Reassess}.

\subsection{Classical Linear Logic}
\label{sec:cll}

Linear Logic, as proposed by Girard~\cite{girard1995linear}, is a refinement to classical logic where the rules of contraction and weakening are limited to the modalities $!$ and $?$. Propositions thus resemble resources that cannot be ignored or copied arbitrarily.

In this work, we use a one-sided sequent calculus version of the multiplicative additive fragment of propositional CLL without units (MALL). Although there exist process translations of full CLL and even first-order CLL, the MALL fragment allows enough expressiveness while keeping the reasoning complexity at a manageable level (MALL is PSPACE-complete whereas full CLL is undecidable~\cite{LINCOLN1992239}). The inference rules for MALL are presented in Fig.~\ref{fig:cll}.

\begin{figure}[htbp]
	\centering
		\begin{tabular}{ccc}
		\vspace*{2mm}
		$\infer[Id]{\tstile A^\bot \sep A}{}$ & &
        $$
		\infer[Cut]{\tstile \Gamma \sep \Delta}{
		\tstile \Gamma \sep C
		&
		\tstile \Delta \sep C^\bot
		}
		$$ \\
		\vspace*{2mm}
		$$
		\infer[\otimes]{\tstile \Gamma \sep \Delta \sep A\otimes B}{
		\tstile \Gamma \sep A
		&
		\tstile \Delta \sep B
		}
		$$ & &
		$$
		\infer[\parr]{\tstile \Gamma \sep (A \otimes B)^\bot}{
		\tstile \Gamma \sep A^\bot \sep B^\bot
		}
		$$ \\
		\vspace*{2mm}
		$$
		\infer[\oplus_L]{\tstile \Gamma \sep A \oplus B}{
		\tstile \Gamma \sep A
		}
		$$ & 
		$$
		\infer[\oplus_R]{\tstile \Gamma \sep A \oplus B}{
		\tstile \Gamma \sep B
		}
		$$ &
		$$
		\infer[\with]{\tstile \Gamma \sep (A\oplus B)^\bot}{
		\tstile \Gamma \sep A^\bot
		&
		\tstile \Gamma \sep B^\bot
		}
		$$ \\
		\end{tabular}\vspace*{-2mm}
			\caption{One-sided sequent calculus versions of the CLL inference rules.}\vspace*{-5mm}	
			\label{fig:cll}
\end{figure}%

In this version of MALL, linear negation ($\cdot ^\bot$) is defined as a syntactic operator with no inference rules, so that both $A$ and $A^\bot$ are considered atomic formulas. The de Morgan style equations in Fig.~\ref{fig:linneg} provide a \emph{syntactic} equivalence of formulas involving negation~\cite{Troelstra92}. This allows us to use syntactically equivalent formulas, such as $A^\bot\parr B^\bot$ and $(A\otimes B)^\bot$ interchangeably. In fact, in the proofs presented in this paper we choose to present formulas containing $\otimes$ and $\oplus$ over their counterparts $\parr$ and $\with$ due to the polarity restrictions we introduce in Section~\ref{sec:spec}.
\begin{figure}[tphb]
	\centering\vspace*{-5mm}
	\begin{tabular}{rclcrclcrcl}
        $(A^\bot)^\bot$ & $\equiv$ & $A$ & $\qquad$ &
        $(A \otimes B)^\bot$ & $\equiv$ & $A^\bot \parr B^\bot$ & $\qquad$ &
		$(A \oplus B)^\bot$ & $\equiv$ & $A^\bot \with B^\bot$ \\
		& & & 
        & $(A \parr B)^\bot$ & $\equiv$ & $A^\bot \otimes B^\bot$ & $\qquad$ &
		$(A \with B)^\bot$ & $\equiv$ & $A^\bot \oplus B^\bot$ \\
	\end{tabular}\vspace*{-2mm}
	\caption{The equations used to define linear negation for MALL.}
	\label{fig:linneg}
\end{figure}%

In the 90s, Abramsky, Bellin and Scott developed the so-called proofs-as-processes paradigm~\cite{abramsky1994proofs,bellin1994}. It involved a correspondence between CLL inference and concurrent processes in the $\pi$-calculus~\cite{milner1999communicating}. They proved that cut-elimination in a CLL proof corresponds to reductions in the $\pi$-calculus translation, which in turn correspond to communication between concurrent processes. As a result, $\pi$-calculus terms constructed via CLL proofs are inherently free of deadlocks. 

The implications of the proofs-as-processes correspondence have been the subject of recent research in concurrent programming by Wadler~\cite{wadler2012propositions}, Pfenning et al.~\cite{acay2016refinements,caires2010,Toninho:2011:DST:2003476.2003499}, Dardha~\cite{dardha2018new,DBLP:journals/corr/DardhaP15} and others. Essentially, each CLL inference step can be translated to an executable workflow, with automatically generated code to appropriately connect the component processes. As a result, the CLL proofs have a direct correspondence to the ``piping'', so to speak, that realises the appropriate resource flow between the available processes, such that it does not introduce deadlocks, accounts for all resources explicitly, and maximizes runtime concurrency. The current paper examines CLL inference and we take the correspondence to deadlock-free processes for granted.

\subsection{Related work}

Diagrammatic languages such as BPMN~\cite{BPMN} are commonly used for the description of workflows in different organisations. However, they typically lack rigour and have limited potential for formal verification~\cite{Szpyrka2012}. Execution languages such as BPEL~\cite{bpel} and process calculi such as Petri Nets~\cite{van1998application} are often used for workflow management in a formal way and our CLL approach could potentially be adapted to work with these. Linear logic has been used in the context of web service composition~\cite{rao2006composition}, but in a way that diverges significantly from the original theory and compromises the validity of the results. Finally, the way the resource flow is managed through our CLL-based processes is reminiscent of monad-like structures such as Haskell's arrows\footnote{https://www.haskell.org/arrows}. One of the key differences is the lack of support for optional resources, which is non-trivial as we show in this paper.

\section{Process Specification}
\label{sec:spec}

Since CLL propositions can naturally represent resources, CLL sequents can be used to represent processes, with each literal representing a type of resource that is involved in that process. These abstract types can have a concrete realisation in the host programming language, from primitive to complicated objects.

Our approach to resource-based composition is to construct CLL specifications of abstract processes based on their inputs (and preconditions) and outputs (and effects), also referred to as IOPEs. This is standard practice in various process formalisms, including WSDL for web services \cite{wsdl}, OWL-S for Semantic Web services \cite{martin2004owl}, PDDL for actions in automated planning~\cite{mcdermott1998pddl}, etc. 

The symmetry of linear negation as shown in Fig.~\ref{fig:linneg} can be used to assign a \emph{polarity} to each CLL connective in order to distinctly specify input and output resources. We choose to treat negated literals, $\parr$, and $\with$ as \textbf{inputs}, and positive literals, $\otimes$, and $\oplus$ as \textbf{outputs}, with the following intuitive interpretation:

\begin{itemize}
\item Multiplicative conjunction (\emph{tensor} $\otimes$) indicates a pair of parallel outputs.

\item Additive disjunction (\emph{plus} $\oplus$) indicates exclusively optional outputs (alternative outputs or exceptions).

\item Multiplicative disjunction (\emph{par} $\parr$) indicates a pair of simultaneous inputs.

\item Additive conjunction (\emph{with} $\with$) indicates exclusively optional input.
\end{itemize}

Based on this, a process can be specified as a CLL sequent consisting of a list of input formulas and a \textbf{single} output formula. In this, the order of the literals does not matter, so long as they obey the polarity restrictions (all but exactly one are negative). In practice, we treat sequents as multisets of literals and manage them using particular multiset reasoning techniques in HOL Light. The description of these techniques is beyond the scope of this paper.

The polarity restrictions imposed on our process specifications match the specification of Laurent's Polarized Linear Logic (LLP)~\cite{laurent2002etude}, and has a proven logical equivalence to the full MALL. Moreover, these restrictions match the programming language paradigm of a \emph{function} that can have multiple input arguments and returns a single (possibly composite) result.

\section{Process Composition}
\label{sec:comp}

Using CLL process specifications as assumptions, we can produce a composite process specification using forward inference. Each of the CLL inference rules represent a logically legal way to manipulate and compose such specifications. 

The axiom $\tstile A \sep A^\bot$ represents the so-called \emph{axiom buffer}, a process that receives a resource of type $A$ and outputs the same resource unaffected.

Unary inference rules, such as the $\oplus_L$ rule, correspond to manipulations of a single process specification. For example, the $\oplus_L$ rule (see Fig.~\ref{fig:cll}) takes a process $P$ specified by $\tstile \Gamma \sep A$, i.e.\ a process with some inputs $\Gamma$ and an output $A$, and produces a process $\tstile \Gamma \sep A \oplus B$, i.e.\ a process with the same inputs $\Gamma$ and output either $A$ or $B$. Note that, in practice, the produced composite process is a transformation of $P$ and thus will always produce $A$ and never $B$.

Binary inference rules, such as the $\otimes$ rule, correspond to binary process composition. The $\otimes$ rule in particular (see Fig.~\ref{fig:cll}) takes a process $P$ specified by $\tstile \Gamma \sep A$ and another process $Q$ specified by $\tstile \Delta \sep B$ and composes them, so that the resulting process $\tstile \Gamma \sep \Delta \sep A\otimes B$ has all their inputs $\Gamma$ and $\Delta$ and a simultaneous output $A \otimes B$. Notably, the $Cut$ rule corresponds to the composition of 2 processes in sequence, where one consumes a resource $A$ given by the other.

Naturally, these manipulations and compositions are primitive and restricted. Constructing meaningful compositions requires several rule applications and, therefore, doing this manually would be a very tedious and impractical task. Our work focuses on creating high level actions that use CLL inference to automatically produce binary process compositions that are correct-by-construction based on the guarantees described above. More specifically, we introduce actions for parallel (\hl{TENSOR}), conditional (\hl{WITH}), and sequential composition (\hl{JOIN}).

Since we are using forward inference, there are infinitely many ways to apply the CLL rules and therefore infinite possible compositions. We are interested in producing compositions that are intuitive for the user. It is practically impossible to produce a formal definition of what these compositions should be. Instead, as explained earlier, we rely on practical experience and user feedback from the various case studies for workflow modelling (see Table~\ref{tbl:projects}). 

Based on this, we have introduced a set of what can be viewed as \emph{unit tests} for our composition actions, which describe the expected and logically valid results of example compositions. As we explore increasingly complex examples in practice, we augment our test set and ensure our algorithms satisfy them. \emph{Selected} unit tests for the \hl{WITH} and \hl{JOIN} actions are shown in Tables \ref{tbl:unit:with} and \ref{tbl:unit:join} respectively. Moreover, as a general principle, our algorithms try to \emph{maximize} resource usage, i.e.\ involve as many resources as possible, and \emph{minimize} the number of rule applications to keep the corresponding process code more compact.

For example, row 3 of Table \ref{tbl:unit:join} indicates that a process with output $A\oplus B$ when composed with a process specified by $\tstile A^\bot \sep B$ should produce a process with output $B$. As we discuss in Section~\ref{sec:action:join:optional}, a different CLL derivation for the same scenario could lead to a process with output $B\oplus B$. This result is unnecessarily more complicated, and its complexity will propagate to all subsequent compositions which will have to deal with 2 options of a type $B$ output. The unit test therefore ensures that the algorithm always leads to a minimal result.

\begin{table}[tpbh]
\centering
\scriptsize
\rowcolors{2}{gray!25}{white}
\begin{tabular}{|c|c|c|}
\hline \proc{P} & \proc{Q} & Result \\\hline\hline
$\tstile X^\bot \sep Z$ & $\tstile Y^\bot \sep Z $ & $\tstile (X\oplus Y)^\bot\sep Z$ \\
$\tstile X^\bot \sep Z$ & $\tstile Y^\bot \sep  W $ & $\tstile (X\oplus Y)^\bot\sep A^\bot\sep B^\bot\sep Z\oplus W$ \\
$\tstile X^\bot \sep A^\bot \sep B^\bot \sep Z$ & $\tstile Y^\bot \sep Z $ & $\tstile (X\oplus Y)^\bot\sep Z\oplus (Z\otimes A\otimes B)$ \\
$\tstile X^\bot \sep A^\bot\sep Z$ & $\tstile Y^\bot \sep B^\bot\sep  W $ & $\tstile (X\oplus Y)^\bot \sep A^\bot\sep B^\bot\sep (Z\otimes B)\oplus (W\otimes A)$ \\
$\tstile X^\bot \sep A^\bot\sep C^\bot\sep Z$ & $\tstile Y^\bot \sep B^\bot\sep  W $ & $\tstile (X\oplus Y)^\bot \sep A^\bot\sep B^\bot\sep C^\bot\sep (Z\otimes B)\oplus (W\otimes A\otimes C)$ \\
$\tstile X^\bot \sep A^\bot\sep C^\bot\sep Z$ & $\tstile Y^\bot \sep B^\bot\sep C^\bot\sep  W $ & $\tstile (X\oplus Y)^\bot \sep A^\bot\sep B^\bot\sep C^\bot\sep (Z\otimes B)\oplus (W\otimes A)$ \\
$\tstile X^\bot \sep A^\bot\sep C^\bot\sep C^\bot\sep Z$ & $\tstile Y^\bot \sep B^\bot\sep C^\bot\sep  W $ & $\tstile (X\oplus Y)^\bot \sep A^\bot\sep B^\bot\sep C^\bot\sep C^\bot\sep (Z\otimes B)\oplus (W\otimes A\otimes C)$ \\
$\tstile X^\bot \sep A\otimes B$ & $\tstile Y^\bot \sep B\otimes A $ & $\tstile (X\oplus Y)^\bot\sep A\otimes B$ \\ 
$\tstile X^\bot \sep A^\bot\sep Z\otimes A $ & $\tstile Y^\bot\sep  Z$ & $\tstile (X\oplus Y)^\bot \sep A^\bot\sep Z \otimes A$ \\
$\tstile X^\bot \sep A^\bot\sep A\otimes Z $ & $\tstile Y^\bot\sep  Z$ & $\tstile (X\oplus Y)^\bot \sep A^\bot\sep A\otimes Z$ \\
$\tstile X^\bot \sep A^\bot\sep Z\oplus(Z\otimes A) $ & $\tstile Y^\bot\sep  Z$ & $\tstile (X\oplus Y)^\bot \sep A^\bot\sep Z\oplus(Z\otimes A)$ \\
\hline
\end{tabular}
\smallskip
\caption{Examples of the expected result of the \hl{WITH} action between $X^\bot$ of a process \proc{P} and $Y^\bot$ of a process \proc{Q}.}
\label{tbl:unit:with}
\end{table}

All our algorithms are implemented within the Higher Order Logic proof tactic system of HOL Light. As a result, the names of some methods have the \hl{\_TAC} suffix, which is conventionally used when naming HOL Light tactics.

\section{Auxiliary Processes}
\label{sec:constr}

During composition, we often need to construct auxiliary processes that manipulate the structure of a CLL type in particular ways. We have identified 2 types of such processes: \emph{buffers} and \emph{filters}.

\paragraph{Buffers:} Similarly to the axiom buffer introduced in the previous section, \emph{composite buffers} (or simply \emph{buffers}) can carry any composite resource without affecting it. This is useful when a process is unable to handle the entire type on its own, and some resources need to be simply \emph{buffered} through. For example, if a process needs to handle a resource of type $A\otimes B$, but only has an input of type $A^\bot$, then $B$ will be handled by a buffer. 

More formally, buffers are processes specified by $\tstile A^\bot \sep A$, where $A$ is arbitrarily complex. Such lemmas are always provable in CLL for any formula $A$. We have introduced an automatic procedure \hl{BUFFER\_TAC} that can accomplish this, but omit the implementation details in the interest of space and in favour of the more interesting composition procedures that follow.

We also introduce the concept of a \emph{parallel buffer}, defined as a process $\tstile A_1^\bot \sep A_2^\bot \sep ... \sep A_n^\bot \sep A_1 \otimes A_2 \otimes ... \otimes A_n$. Such buffers are useful when composing processes with an optional output (see  Section~\ref{sec:action:join:optional}). Their construction can also be easily automated with a decision procedure we call \hl{PARBUF\_TAC}.

\paragraph{Filters:} Often during process composition by proof, resources need to match exactly for the proof to proceed. In some cases, composite resources may not match exactly, but may be manipulated using the CLL inference rules so that they end up matching. For example, the term $A \otimes B$ does not directly match $B \otimes A$. However, both terms intuitively represent resources $A$ and $B$ in parallel. This intuition is reflected formally to the commutativity property of $\otimes$, which is easily provable in CLL: $\tstile (A \otimes B)^\bot \sep B \otimes A$. We can then use the $Cut$ rule with this property to convert an output of type $A\otimes B$ to $B\otimes A$ (similarly for inputs). 

We call such lemmas that are useful for converting CLL types to logically equivalent ones, \emph{filters}. In essence, a filter is any provable CLL lemma that preserves our polarity restrictions. We prove such lemmas automatically using the proof strategies developed by Tammet~\cite{Tammet1994}. 
We call such lemmas that are useful for converting CLL types to logically equivalent ones, \emph{filters}. In essence, a filter is any provable CLL lemma that preserves our polarity restrictions. We prove such lemmas automatically using the proof strategies developed by Tammet~\cite{Tammet1994}. 

We give some examples of how filters are used to match terms as we go through them below. However, as a general rule the reader may assume that, for the remainder of this paper, by ``equal'' or ``matching'' terms we refer to terms that are equal modulo the use of filters.

A main consequence of this is that our algorithms often attempt to match literals that do not match. For example, the attempt to compose $\tstile A^\bot\sep B$ in sequence with $\tstile C^\bot\sep D^\bot\sep E$ would generate and try to prove 2 false conjectures $\tstile B^\bot\sep C$ and $\tstile B^\bot\sep D$ in an effort to match the output $B$ with any of the 2 inputs $C^\bot$ and $D^\bot$ before failing\footnote{In practice, the user will have to select a matching input to attempt such a composition (see Section~\ref{sec:action:join}).}. This highlights the need for an efficient proof procedure for filters, with an emphasis on early failure.

\begin{table}[tpbh]
\centering
\scriptsize
\rowcolors{2}{gray!25}{white}
\begin{tabular}{|c|c|c|c|c|}
\hline \proc{P} & Pr. & \proc{Q} & Selected Input & Result \\\hline\hline
$\tstile X^\bot \sep A$ & $$ & $\tstile A^\bot \sep Y $ & $A^\bot$ & $\tstile X^\bot\sep Y$ \\
$\tstile X^\bot \sep A\otimes B$ & $L$ & $\tstile A^\bot \sep Y $ & $A^\bot$ & $\tstile X^\bot\sep Y \otimes B$ \\
$\tstile X^\bot \sep A\oplus B$ & $L$ & $\tstile A^\bot \sep B $ & $A^\bot$ & $\tstile X^\bot\sep B$ \\
$\tstile X^\bot \sep A\otimes B\otimes C$ & $L$ & $\tstile A^\bot \sep Y $ & $A^\bot$ & $\tstile X^\bot\sep Y \otimes B\otimes C$ \\
$\tstile X^\bot \sep A\oplus B$ & $L$ & $\tstile A^\bot \sep C^\bot \sep Y $ & $A^\bot$ & $\tstile X^\bot\sep C^\bot\sep Y \oplus (C\otimes B)$ \\
$\tstile X^\bot \sep A\oplus B$ & $R$ & $\tstile B^\bot \sep C^\bot \sep Y $ & $B^\bot$ & $\tstile X^\bot\sep C^\bot\sep (C\otimes A)\oplus Y$ \\
$\tstile X^\bot \sep A\oplus B$ & $L$ & $\tstile (B\oplus A)^\bot \sep Y $ & $(B\oplus A)^\bot$ & $\tstile X^\bot\sep Y$ \\
$\tstile X^\bot \sep A\oplus (B\otimes C)$ & $L$ & $\tstile (B\oplus A)^\bot \sep Y $ & $(B\oplus A)^\bot$ & $\tstile X^\bot\sep Y\oplus(B\otimes C)$ \\
$\tstile X^\bot \sep A\oplus (B\otimes C)$ & $RL$ & $\tstile (B\oplus A)^\bot \sep Y $ & $(B\oplus A)^\bot$ & $\tstile X^\bot\sep A\oplus(Y\otimes C)$ \\
$\tstile X^\bot \sep A\oplus B$ & $L$ & $\tstile (C\oplus A\oplus D)^\bot \sep Y $ & $(C\oplus A\oplus D)^\bot$ & $\tstile X^\bot\sep Y\oplus B$ \\
$\tstile X^\bot \sep C\oplus (A\otimes B)$ & $L$ & $\tstile C^\bot \sep A\otimes B $ & $C^\bot$ & $\tstile X^\bot\sep A\otimes B$ \\
$\tstile X^\bot \sep C\oplus (A\otimes B)$ & $L$ & $\tstile C^\bot \sep B\otimes A $ & $C^\bot$ & $\tstile X^\bot\sep B\otimes A$ \\
$\tstile X^\bot \sep C\oplus (A\otimes (B\oplus D))$ & $L$ & $\tstile C^\bot \sep (B\oplus D)\otimes A $ & $C^\bot$ & $\tstile X^\bot\sep (B\oplus D)\otimes A $ \\
$\tstile X^\bot \sep C\oplus (A\otimes B)$ & $L$ & $\tstile C^\bot \sep Y\oplus (B\otimes A)$ & $C^\bot$ & $\tstile X^\bot\sep Y\oplus (B\otimes A) $ \\
$\tstile X^\bot \sep C\oplus (A\otimes B)$ & $L$ & $\tstile C^\bot \sep (B\otimes A)\oplus Y$ & $C^\bot$ & $\tstile X^\bot\sep  (B\otimes A)\oplus Y $ \\
$\tstile X^\bot \sep (A\otimes B)\oplus C$ & $R$ & $\tstile C^\bot \sep Y\oplus (B\otimes A)$ & $C^\bot$ & $\tstile X^\bot\sep Y\oplus (B\otimes A) $ \\
$\tstile X^\bot \sep (A\otimes B)\oplus C$ & $R$ & $\tstile C^\bot \sep (B\otimes A)\oplus Y$ & $C^\bot$ & $\tstile X^\bot\sep  (B\otimes A)\oplus Y $ \\
$\tstile X^\bot \sep C\oplus (A\otimes B)$ & $L$ & $\tstile C^\bot \sep Y\oplus (B\otimes A)$ & $C^\bot$ & $\tstile X^\bot\sep Y\oplus (B\otimes A) $ \\
\hline
\end{tabular}
\smallskip
\caption{Examples of the expected result of the \hl{JOIN} action between a process \proc{P} and a process \proc{Q}. Column \emph{Pr.}\ gives the \emph{priority} parameter (see Section~\ref{sec:algo}).}\label{tbl:unit:join}
\end{table}

\section{Parallel Composition - The \hl{TENSOR} Action}

The \hl{TENSOR} action corresponds to the parallel composition of two processes so that their outputs are provided in parallel. It trivially relies on the tensor ($\otimes$) inference rule. Assuming 2 processes, $\tstile A^\bot\sep C^\bot\sep D$ and $\tstile B^\bot\sep E$, the \hl{TENSOR} action will perform the following composition:
{\footnotesize
\[
\infer[\otimes]{\tstile A^\bot\sep B^\bot\sep C^\bot\sep D\otimes E}
{\tstile A^\bot\sep C^\bot\sep D
&
\tstile B^\bot\sep E}
\]}%
\section{Conditional Composition - The \hl{WITH} Action}
\label{sec:action:with}

The \hl{WITH} action corresponds to the \emph{conditional} composition of two processes. This type of composition is useful in cases where each of the components of an optional output of a process needs to be handled by a different receiving process. 

For example, assume a process \proc{S} has an optional output $A \oplus C$ where $C$ is an exception. We want $A$ to be handled by some process \proc{P}, for example specified by $\tstile A^\bot \sep B^\bot \sep X$, while another process \proc{Q} specified by $\tstile C^\bot \sep Y$ plays the role of the exception handler for exception $C$. For this to happen, we need to compose \proc{P} and \proc{Q} together using the \hl{WITH} action so that we can cnostruct an input that matches the output type $A \oplus C$ from \proc{S}. This composition can be viewed as the construction of an \emph{if-then} statement where if $A$ is provided then \proc{P} will be executed (assuming $B$ is also provided), and if $C$ is provided then \proc{Q} will be executed in a mutually exclusive choice. The generated proof tree for this particular example is the following: 

\begin{equation}
\label{eq:action:with}
\infer [\with]{\memph{\tstile (A \oplus C)^\bot \sep B^\bot \sep X\oplus(Y\otimes B)}}{
	\infer [\oplus L]{\tstile A^\bot \sep B^\bot \sep X\oplus(Y\otimes B)}{
		\infer[\mathproc{P}]{\memph{\tstile A^\bot \sep B^\bot \sep X}}{}
	}
	&
	\infer [\oplus R]{\tstile C^\bot \sep B^\bot \sep X\oplus(Y\otimes B)}{
		\infer[\otimes]{\tstile C^\bot \sep B^\bot \sep Y\otimes B}{
			\infer[\mathproc{Q}]{\memph{\tstile C^\bot \sep Y}}{}
			&
			\infer[Id]{\tstile B^\bot \sep B}{}
		}
	}	
}
\end{equation}

The \hl{WITH} action fundamentally relies on the $\with$ rule of CLL. The following derivation allows us to compose 2 processes that also have different outputs $X$ and $Y$:

\begin{equation}
\label{eq:withproc}
\infer[\with]{\tstile \Gamma \sep (A \oplus C)^\bot \sep X \oplus Y}{
			\infer[\oplus L]{\tstile \Gamma \sep A^\bot \sep X\oplus Y}{
				\tstile \Gamma \sep A^\bot \sep X
			}
			&
			\infer[\oplus R]{\tstile \Gamma \sep C^\bot \sep X\oplus Y}{
				\tstile \Gamma \sep C^\bot \sep Y
			}
		}
\end{equation}

The particularity of the $\with$ rule is that the context $\Gamma$, i.e.\ all the inputs except the ones involved in the \hl{WITH} action, must be the same for both the involved processes. In practice, this means we need to account for unused inputs. In the example above, \proc{P} apart from input $A^\bot$ has another input $B^\bot$ which is missing from \proc{Q}. In the conditional composition of \proc{P} and \proc{Q}, if exception $C$ occurs, the provided $B$ will not be consumed since \proc{P} will not be invoked. In this case, we use a buffer to let $B$ pass through together with the output $Y$ of \proc{Q}.

More generally, in order to apply the $\with$ rule to 2 processes $\proc{P}$ and $\proc{Q}$, we need to minimally adjust their contexts $\Gamma_{P}$ and $\Gamma_{Q}$ (i.e.\ their respective multisets of inputs excluding the ones that will be used in the rule) so that they end up being the same $\Gamma = \Gamma_{P} \union \Gamma_{Q}$. By \quot{minimal} adjustment we mean that we only add the inputs that are \quot{missing} from either side, i.e.\ the multiset $\Delta_{P} = \Gamma_{Q} \setminus \Gamma_{P}$ for \proc{P} and $\Delta_{Q} = \Gamma_{P} \setminus \Gamma_{Q}$ for \proc{Q}, and no more.

In the previous example in \eqref{eq:action:with}, exculding the inputs $A^\bot$ and $C^\bot$ used in the rule, we obtain $\Delta_Q = \Gamma_{P} \setminus \Gamma_{Q} = \{B^\bot\} \setminus \{\} = \{B^\bot\}$. We then construct a parallel buffer (see Section~\ref{sec:constr}) of type $\otimes\Delta_{Q}^\bot$\footnote{$\otimes\{a_1,..., a_n\}^\bot = a_{1}^\bot \otimes ... \otimes a_{n}^\bot$} (converting all inputs in $\Delta_{Q}$ to an output; in this example only one input) using \hl{PARBUF\_TAC}. In the example, this is an atomic $B$ buffer. The parallel composition between this buffer and \proc{Q} results in the process $\tstile \Gamma_{Q} \sep \Delta_{Q} \sep C^\bot \sep Y \otimes (\otimes\Delta_{Q}^\bot)$. The same calculation for \proc{P} yields $\Delta_{P} = \emptyset$ so no change is required for \proc{P}.

Since $\Gamma_{P} \munion \Delta_{P} = \Gamma_{Q} \munion \Delta_{Q} = \Gamma$ (where $\munion$ denotes multiset union), the $\with$ rule is now applicable and derivation \eqref{eq:withproc} yields the following process:

\begin{equation}
	\tstile \Gamma \sep (A \oplus C)^\bot \sep \left( X \otimes (\otimes\Delta_{P}^\bot) \right) \oplus \left( Y \otimes (\otimes\Delta_{Q}^\bot) \right)
\label{eq:cond:general}
\end{equation}

The output $Y$ of $Q$ has now been paired with the buffered resources $\Delta_{Q}$.

Finally, we consider the special case where the following holds:
\begin{equation}
\label{eq:with:special}
\left( X \otimes (\otimes\Delta_{P}^\bot) \right) = \left( Y \otimes (\otimes\Delta_{Q}^\bot) \right) = G
\end{equation}

In this case, the output of the composition in \eqref{eq:cond:general} will be $G\oplus G$. Instead we can apply the $\with$ directly without derivation \eqref{eq:withproc}, yielding the simpler output $G$. 

Note that, as discussed in Section~\ref{sec:constr}, \eqref{eq:with:special} above does not strictly require equality. The special case can also be applied if we can prove and use the filter {\footnotesize$\tstile \left( X \otimes (\otimes\Delta_{P}^\bot) \right)^\bot \sep \left( Y \otimes (\otimes\Delta_{P}^\bot) \right)$}.

These results and the complexity underlying their construction demonstrate the non-trivial effort needed to adhere to CLL's systematic management of resources and, more specifically, its systematic accounting of unused resources. These properties, however, are essential guarantees of correct resource management offered by construction in our process compositions.

\section{Sequential Composition - The \hl{JOIN} Action}
\label{sec:action:join}

The \hl{JOIN} action reflects the connection of two processes in sequence, i.e.\ where (some of) the outputs of a process are connected to (some of) the corresponding inputs of another. More generally, we want to compose a process \proc{P} with specification $\tstile \Gamma \sep X$, i.e.\ with some (multiset of) inputs $\Gamma$ and output $X$ in sequence with a process \proc{Q} with specification $\tstile \Delta \sep C^\bot \sep Y$, i.e.\ with an input $C^\bot$, output $Y$, and (possibly) more inputs in context $\Delta$. We also assume the user selects a subterm $A$ of $X$ in \proc{P} and a matching subterm $A$ of the input $C^\bot$ in \proc{Q}.

The strategy of the algorithm behind the \hl{JOIN} action is to construct a new input for \proc{Q} based on the chosen $C^\bot$ such that it directly matches the output $X$ of \proc{P} (and prioritizing the output selection $A$). This will enable the application of the $Cut$ rule, which requires the cut literal to match exactly. In what follows, we present how different cases for $X$ are handled.

\subsection{Atomic or Matching Output}
\label{sec:action:join:atomic}

If $X$ is atomic, a straighforward use of the $Cut$ rule is sufficient to connect the two processes. For example, the \hl{JOIN} action between $\tstile A^\bot \sep B^\bot \sep X$ and $\tstile X^\bot \sep Z$ results in the following proof:

\begin{equation}
\nonumber
\infer[Cut]{\memph{\tstile A^\bot \sep B^\bot \sep Z}}{
	\infer[\mathproc{P}]{\memph{\tstile A^\bot \sep B^\bot,\ X}}{}
	&
	\infer[\mathproc{Q}]{\memph{\tstile X^\bot \sep Z}}{}
}%
\end{equation}

The same approach can be applied more generally for any non-atomic $X$ as long as a matching input of type $X^\bot$ (including via filtering) is selected in \proc{Q}.

\subsection{Parallel Output}
\label{sec:action:join:parallel}

If $X$ is a parallel output, such as $B \otimes C$, we need to manipulate process \proc{Q} so that it can receive an input of type $(B \otimes C)^\bot$.

If \proc{Q} has both inputs $B^\bot$ and $C^\bot$, then we can use the $\parr$ rule to combine them. For example, the generated proof tree of the \hl{JOIN} action between $\tstile A^\bot \sep D^\bot \sep B \otimes C$ and $\tstile B^\bot \sep C^\bot \sep E^\bot \sep Y$ is the following:
{\footnotesize
\[
\infer[Cut]{\tstile A^\bot \sep D^\bot \sep E^\bot \sep Y}{
	\infer[\mathproc{P}]{\tstile A^\bot \sep D^\bot \sep B \otimes C}{}
	&
	\infer[\parr]{\tstile (B\otimes C)^\bot \sep E^\bot \sep Y}{
		\infer[\mathproc{Q}]{\tstile B^\bot \sep C^\bot \sep E^\bot \sep Y}{}
	}
}\]}%

As previously mentioned, the \hl{JOIN} action attempts to connect the output of \proc{P} to \proc{Q} maximally, i.e.\ both $B$ and $C$, regardless of the user choice. The user may, however, want to only connect one of the two resources. We have currently implemented this approach as it is the most commonly used in practice, but are investigating ways to enable better control by the user.
 
If \proc{Q} has only one of the two inputs, for example $B^\bot$, i.e.\ \proc{Q} is of the form $\tstile \Delta \sep B^\bot \sep Y$ and $C^\bot \not \in \Delta$, then $C$ must be buffered. In this case, we use the following derivation:
\begin{equation}
\label{eq:join:par:buf}
\infer[\parr]{\tstile \Delta \sep (B\otimes C)^\bot \sep Y \otimes C}{
	\infer[\otimes]{\tstile \Delta \sep B^\bot \sep C^\bot \sep Y \otimes C}{
      \infer[\mathproc{Q}]{\tstile \Delta \sep B^\bot \sep Y}{}
      &
      \infer*{\tstile C^\bot \sep C}{\mathhl{BUFFER\_TAC}}
  }
}
\end{equation}

We use \hl{BUFFER\_TAC} from Section~\ref{sec:constr} to prove the buffer of $C$.

Depending on the use of the $\otimes$ rule in \eqref{eq:join:par:buf}, the resulting output could be either $Y\otimes C$ or $C\otimes Y$. We generally try to match the form of \proc{P}'s output, so in this case we would choose $Y\otimes C$ to match $B\otimes C$. Our algorithm keeps track of this orientation through the \hl{orient} parameter (see Section~\ref{sec:algo}).

\subsection{Optional Output}
\label{sec:action:join:optional}

If $X$ is an optional output, such as $B \oplus C$, then we need to manipulate process \proc{Q} to synthesize an input $(B \oplus C)^\bot$. Assume \proc{Q} can handle $B$ (symmetrically for $C$) and thus has specification $\tstile \Delta \sep B^\bot \sep Y$. We construct a parallel buffer (using \hl{PARBUF\_TAC}, see Section~\ref{sec:constr}) of type $(\otimes\Delta^\bot) \otimes C$ (converting all inputs in $\Delta$ to outputs). We then apply derivation \eqref{eq:withproc} as follows:
{\footnotesize
\begin{equation}
\infer[\with]{\tstile \Delta \sep (B \oplus C)^\bot \sep Y \oplus \left((\otimes\Delta^\bot) \otimes C\right)}{
	\infer[\oplus L]{\tstile \Delta \sep B^\bot \sep Y \oplus \left((\otimes\Delta^\bot) \otimes C\right)}{
		\infer[\mathproc{Q}]{\tstile \Delta \sep B^\bot \sep Y}{}
	}
	&
	\infer[\oplus R]{\tstile \Delta \sep C^\bot \sep Y \oplus \left((\otimes\Delta^\bot) \otimes C\right)}{
	\infer*{\tstile \Delta \sep C^\bot \sep (\otimes\Delta^\bot) \otimes C}{\mathhl{PARBUF\_TAC}}
	}
}
\label{eq:join:optional:output:proof}
\end{equation}
}%

Similarly to the \hl{WITH} action, the particular structure of the $\with$ rule ensures the systematic management of unused resources. In the example above, if $C$ is received then \proc{Q} will never be executed. As a result, any resources in $\Delta$ will remain unused and need to be buffered together with $C$. This is the reason behind the type $(\otimes\Delta^\bot) \otimes C$ of the constructed buffer (as opposed to plainly using type $C$).

The proof tree of an example of the \hl{JOIN} action between process \proc{P} specified by $\tstile A^\bot \sep D^\bot \sep B \oplus C$ and process \proc{Q} specified by $\tstile B^\bot \sep E^\bot \sep Y$ is shown below:
{\footnotesize
\[
\infer[Cut]{\memph{\tstile A^\bot \sep D^\bot \sep E^\bot \sep Y\oplus (C\otimes E)}}{
	\infer[\mathproc{P}]{\memph{\tstile A^\bot \sep D^\bot \sep B \oplus C}}{}
	&
	\infer[\with]{\tstile (B\oplus C)^\bot \sep E^\bot \sep Y\oplus (C\otimes E)}{
		\infer[\oplus L]{\tstile B^\bot \sep E^\bot \sep Y\oplus (C\otimes E)}{
			\infer[\mathproc{Q}]{\tstile B^\bot \sep E^\bot \sep Y}{}
		}
		&
		\infer[\oplus R]{\tstile C^\bot \sep E^\bot \sep Y\oplus (C\otimes E)}{
			\infer[\otimes]{\tstile C^\bot \sep E^\bot \sep C\otimes E}{
				\infer[Id]{\tstile C^\bot \sep C}{}
				&
				\infer[Id]{\tstile E^\bot \sep E}{}
			}
		}
	}
}
\]
}
It is interesting to consider a couple of special cases.

\emph{Case 1:} If $\tstile \Delta \sep C^\bot \sep Y$ is a parallel buffer, \eqref{eq:join:optional:output:proof} can be simplified as follows:

\begin{equation}
\label{eq:join:optional:output:proof:simplified:1}
\infer[\with]{\tstile \Delta \sep (B \oplus C)^\bot \sep Y}{
	\infer[\mathproc{Q}]{\tstile \Delta \sep B^\bot \sep Y}{}
	&
	\infer*{\tstile \Delta \sep C^\bot \sep Y}{\mathhl{PARBUF\_TAC}}
}
\end{equation}

This may occur, for example, if $\Delta = \emptyset$ and $Y = C$. Such cases arise in processes used to recover from an exception. For instance, a recovery process $\tstile Exception^\bot\sep Resource$ can convert an output $Resource\oplus Exception$ to simply $Resource$ (which either was there in the first place, or was produced through the recovery process).

\emph{Case 2:} If $Y = D \oplus E$ for some $D$ and $E$ such that $\tstile \Delta \sep C^\bot \sep D$ (or symmetrically $\tstile \Delta \sep C^\bot \sep E$) is a parallel buffer, then we can apply the following derivation:
\begin{equation}
\label{eq:join:optional:output:proof:simplified:2}
\infer[\with]{\tstile \Delta \sep (B \oplus C)^\bot \sep D \oplus E}{
	\infer[\mathproc{Q}]{\tstile \Delta \sep B^\bot \sep D \oplus E}{}
	&
    \infer[\oplus L]{\tstile \Delta \sep C^\bot \sep D \oplus E}{
	\infer*{\tstile \Delta \sep C^\bot \sep D}{\mathhl{PARBUF\_TAC}}
}}
\end{equation}

This may occur, for example, if $\Delta = \emptyset$ and $Y = C \oplus E$. The recovery process above may itself throw an exception: $\tstile Exception^\bot\sep Resource\oplus Failed$. This will convert output $Resource\oplus Exception$ to $Resource\oplus Failed$ (either we had the $Resource$ from the beginning, or we recovered and still got a $Resource$, or the recovery process failed) instead of $(Resource\oplus Failed) \oplus Resource$. 

\begin{table}[tpb]
\centering
\begin{tabular}{|c|c|c|c|}
\hline Target & Priority & \proc{Q} & Result of \inputtac \\\hline\hline
$X = \mathbf{A} \otimes (A\oplus B)$ & $Left$ & $\ \tstile A^\bot \sep \mathbf{Y}\ $ & $\ \tstile X^\bot\sep \mathbf{Y}\otimes (A\oplus B)\ $ \\
$X = A \otimes (\mathbf{A}\oplus B)$ & $Right;\ Left$ & $\ \tstile A^\bot \sep \mathbf{Y}\ $ & $\ \tstile X^\bot\sep A\otimes (\mathbf{Y}\oplus B)\ $ \\\hline
$X = \mathbf{A} \oplus (B\otimes C)$ & $Left$ & $\ \tstile (B \oplus A)^\bot \sep \mathbf{Y}\ $ & $\ \tstile X^\bot\sep \mathbf{Y}\oplus (B\otimes C)\ $ \\
$X = A \oplus (\mathbf{B}\otimes C)$ & $Right;\ Left$ & $\ \tstile (B \oplus A)^\bot \sep \mathbf{Y}\ $ & $\tstile X^\bot\sep A\oplus (\mathbf{Y}\otimes C)\ $ \\\hline
\end{tabular}
\smallskip
\caption{Examples of how the \hl{priority} parameter can affect the behaviour of \inputtac. The selected subterms and the output of \proc{Q} are highlighted in bold.}
\label{tbl:priority}
\end{table}

\subsection{Putting It All Together}
\label{sec:algo}

In the general case, the output $X$ of \proc{P} can be a complex combination of multiple parallel and optional outputs. For that reason, we apply the above proof strategies in a recursive, bottom-up way, prioritizing the user selections. We call the algorithm that produces the appropriate input $X^\bot$ (or equivalent) from \proc{Q} ``\inputtac'' and it has the following arguments (see Algorithm~\ref{alg:join}):

\begin{itemize}
\item \textbf{sel}: optional term corresponding to the user selected input $C^\bot$ of \proc{Q}.
\item \textbf{priority}: a list representing the path of the user selected subterm $A$ in the syntax tree of the output $X$ of \proc{P}. For example, if the user selects $B$ in the output $(A \otimes B)\oplus C$, the priority is $\lbrack Left;\ Right\rbrack$.
\item \textbf{orient}: our latest path (left or right) in the syntax tree of $X$ so that we add the corresponding buffers on the same side (see Section~\ref{sec:action:join:parallel}).
\item \textbf{inputs}: a list of inputs of \proc{Q}. We remove used inputs from this to avoid reuse.
\item \textbf{target}: the input term we are trying to construct. This is initially set to $X$, but may take values that are subterms of $X$ in recursive calls.
\item \textbf{proc}: the CLL specification of \proc{Q} as it evolves.
\end{itemize}

The \hl{priority} parameter is useful when more than one subterms of the output either (a) are the same or (b) have the same matching input in \proc{Q}. Table~\ref{tbl:priority} shows examples of how different priorities change the result of \inputtac.

\begin{algorithm}[!htb]
\algrenewcommand\algorithmicindent{0.9em}
  \footnotesize
  \caption{\footnotesize Derives a new process specification from the given ``\hl{proc}'' such that it includes an input of type ``\hl{target}''.\label{alg:join}}
   \begin{algorithmic}[1]
   \Function{\inputtac}{\hl{sel, priority, orient, inputs, target, proc}}
   \State Try to match \hl{target} with \hl{sel} (if provided) or one of the \hl{inputs}
   \If{it matches} \Return \hl{proc}
    \medskip
   \ElsIf{\hl{target} is atomic}
   	\If{\hl{priority} $\neq$ \hl{None}} fail \Comment{we couldn't match the user selected output}
    \Else \ Create a \hl{target} buffer using \eqref{eq:join:par:buf} depending on \hl{orient}
    \EndIf
   \medskip
   \ElsIf{\hl{target} is $L\otimes R$} 
   	\If{\hl{priority} = \hl{Left}}
    	\State \hl{proc'} = \Call{\inputtac}{\hl{sel, tail(priority), orient, inputs, $L$, proc}}
        \State \hl{proc} = \Call{\inputtac}{\hl{None, None, Right, inputs - \{$L$\}, $R$, proc'}}
    \Else
    	\State \hl{proc'} = \Call{\inputtac}{\hl{sel, tail(priority), orient, inputs, $R$, proc}}
        \State \hl{proc} = \Call{\inputtac}{\hl{None, None, Left, inputs - \{$R$\}, $L$, proc'}}
    \EndIf
    \State Use the $\parr$ rule to create the $(L\otimes R)^\bot$ input
   \medskip
   \ElsIf{\hl{target} is $L\oplus R$} 
   	\If{\hl{priority} = \hl{Left}}
    	\State \hl{proc} = \Call{\inputtac}{\hl{sel, tail(priority), orient, inputs, $L$, proc}}
        \State Try derivation \eqref{eq:join:optional:output:proof:simplified:1} \textbf{orElse} Try derivation \eqref{eq:join:optional:output:proof:simplified:2} \textbf{orElse} Use derivation \eqref{eq:join:optional:output:proof} 
    \ElsIf{\hl{priority} = \hl{Right}}
    	\State \hl{proc} = \Call{\inputtac}{\hl{sel, tail(priority), orient, inputs, $R$, proc}}
         \State Try derivation \eqref{eq:join:optional:output:proof:simplified:1} \textbf{orElse} Try derivation \eqref{eq:join:optional:output:proof:simplified:2} \textbf{orElse} Use derivation \eqref{eq:join:optional:output:proof}
    \Else
    	\State Try as if \hl{priority} = \hl{Left} \textbf{orElse} Try as if \hl{priority} = \hl{Right} 
        \State \textbf{else} Create a \hl{target} buffer using \eqref{eq:join:par:buf} depending on \hl{orient} 
    \EndIf
   \EndIf 
   \State \Return \hl{proc}
   \EndFunction
   \end{algorithmic}%
\end{algorithm}

\section{Conclusion}

CLL's inherent properties make it an ideal language to reason about resources. CLL sequents (under polarity restrictions) can be viewed as resource-based specifications of processes. The CLL inference rules then describe the logically legal, but primitive ways to manipulate and compose such processes. 

We presented algorithms that allow intuitive composition in parallel, conditionally, and in sequence. We call these composition actions \hl{TENSOR}, \hl{WITH}, and \hl{JOIN} respectively, and they are implemented in HOL Light. We analysed how each action functions in different cases and examples. 

As a result of the rigorous usage of CLL inference rules, the constructed compositions have guaranteed resource accounting, so that no resources disappear or are created out of nowhere. The proofs-as-processes paradigm and its recent evolutions allow the extraction of process calculus terms from these proofs, for concurrent and guaranteed deadlock-free execution.

In the future, we intend to work towards relaxing identified limitations along 2 main lines: (a) functionality, by incorporating and dealing with increasingly more complex specifications including those requiring formulation of more complex filters, and (b) expressiveness, by extending the fragment of CLL we are using while keeping a balance in terms of efficiency.

Through this work, it is made obvious that intuitive process compositions in CLL require complex applications of a large number of inference rules. Our algorithms automate the appropriate deductions and alleviate this burden from the user. We have tied these with the diagrammatic interface of \hl{WorkflowFM}~\cite{workflowfm}, so that the user is not required to know or understand CLL or theorem proving, but merely sees inputs and outputs represented graphically. They can then obtain intuitive process compositions with the aforementioned correctness guarantees with a few simple clicks.

\section*{Acknowledgements}

This work was supported by the ``DigiFlow: Digitizing Industrial Workflow, Monitoring and Optimization'' Innovation Activity funded by EIT Digital. We would like to thank the attendants of the LOPSTR conference, 4-6 September 2018, in Frankfurt, Germany for their insightful comments that helped improve this paper.

%
%
\bibliographystyle{splncs03}
\bibliography{main}

\end{document}